\documentclass[preprint, showpacs,amsmath,amssymb,prb]{revtex4}
\usepackage{graphicx}

\begin{document}

\title{Ultrafast optical switching of three-dimensional Si inverse opal photonic band gap crystals}

\author{Tijmen G. Euser}
\affiliation{FOM Institute for Atomic and Molecular Physics,
Kruislaan 407, 1098 SJ Amsterdam, The Netherlands}

\author{Hong Wei}
\affiliation{Department of Chemical Engineering \& Materials
Science, University of Minnesota, Minneapolis, Minnesota 55455, USA}

\author{Jeroen Kalkman}
\affiliation{FOM Institute for Atomic and Molecular Physics,
Kruislaan 407, 1098 SJ Amsterdam, The Netherlands }

\author{Yoonho Jun}
\affiliation{Department of Chemical Engineering \& Materials
Science, University of Minnesota, Minneapolis, Minnesota 55455, USA }

\author{Albert Polman}
\affiliation{ FOM Institute for Atomic and Molecular Physics,
Kruislaan 407, 1098 SJ Amsterdam, The Netherlands}

\author{David J. Norris}
\affiliation{ Department of Chemical Engineering \& Materials
Science, University of Minnesota, Minneapolis, Minnesota 55455, USA   }

\author{Willem L. Vos}
\affiliation{FOM Institute for Atomic and Molecular Physics,
Kruislaan 407, 1098 SJ Amsterdam, The Netherlands}
\affiliation{Complex Photonic Systems (COPS), MESA$^+$ Research
Institute, University of Twente, The Netherlands}

\email{w.l.vos@amolf.nl} \homepage{www.photonicbandgaps.com}


\begin{abstract}
We present ultrafast optical switching experiments on 3D
photonic band gap crystals. Switching the Si inverse opal
is achieved by optically exciting free carriers by a
two-photon process. We probe reflectivity in the frequency
range of second order Bragg diffraction where the photonic
band gap is predicted.  We find good experimental switching
conditions for free-carrier plasma frequencies between 0.3
and 0.7 times the optical frequency $\omega$: we thus observe a
large frequency shift of up to $\Delta\omega/\omega$= 1.5$\%$ of
all spectral features including the peak that corresponds to the
photonic band gap. We deduce a corresponding large refractive index
change of $\Delta n'_{Si}/n'_{Si}=$ 2.0$\%$, where $n'_{Si}$ is the
refractive index of the silicon backbone of the crystal.
The induced absorption length that is longer than the
sample thickness. We observe a fast decay time of 21 ps,
which implies that switching could potentially be repeated
at GHz rates. Such a high switching rate is relevant to future switching and
modulation applications.
\end{abstract}

\maketitle

\section{INTRODUCTION}
\label{introduction} Currently, many efforts are devoted to an intricate class of three-dimensional metamaterials known as photonic crystals.\cite{Crete01}
Spatially periodic variations of the refractive index commensurate with optical
wavelengths cause the photon dispersion relations to organize in bands, analogous to electron
bands in solids. Generally, frequency windows known as stop gaps
appear in which modes are forbidden for specific wave vectors.
Experimentally, stop gaps appear as peaks in reflectivity spectra.
The strong dispersion in photonic crystals can be used to control
the propagation direction of light. Fundamental interest in 3D
photonic crystals is spurred by the possibility of a photonic band
gap, a frequency range for which no modes exist at all. Tailoring of
the photonic density of states by a photonic crystal allows one to
control fundamental atom-radiation interactions in solid-state
environments.\cite{Yablonovitch87,Lodahl04} In this context the
hallmark of a photonic band gap is the eagerly awaited inhibition of
spontaneous emission due to a vanishing density of states.\cite{Yablonovitch87}

Exciting prospects arise when 3D photonic band gap crystals are
switched on ultrafast timescales. First of all, switching the
directional properties of photonic crystals leads to fast changes in
the reflectivity.  Ultrafast control of the propagation of light, which was
demonstrated in 2D photonic crystals by Ref. \cite{Leonard02},  is
essential to applications in active photonic integrated circuits.\cite{Nakamura04} Secondly, switching would also allow the capture
or release of photons from photonic band gap cavities,\cite{Johnson02} which is relevant to solid-state slow-light
schemes.\cite{Yanik04} Thirdly, switching 3D photonic band gap
crystals provides dynamic control over the density of states that
would allow the switching-on or -off of light sources in the band
gap.\cite{Johnson02}

An optical switching experiment requires a switching
magnitude as large as possible, ultrafast time-scales, as low as
possible induced absorption, as well as good spatial homogeneity
of the index change.\cite{Johnson02} In our experiment we change the refractive index
of the semiconductor backbone of the crystal by optically exciting a
free-carrier plasma. The refractive index of the excited crystal is
well described by the Drude model,\cite{Sokolowski00} in which the plasma frequency
$\omega_p$ is proportional to the density of excited carriers.  By
carefully choosing the amount of excited carriers, and thus the
plasma frequency, large changes in refractive index are feasible,
while the induced absorption is predicted to remain
small.\cite{Johnson02} For Si, good experimental conditions are
found for free-carrier plasma frequencies around $\omega_p$=
0.3-0.7$\times \omega_{probe}$, where $\omega_{probe}$ is the frequency of the probe light.
The spatial homogeneity of the excited carrier plasma can be optimized by choosing a two-photon
excitation mechanism rather than a linear process.\cite{Euser05} In
Si photonic crystals, optimum spatial homogeneity is obtained for
pump frequencies near the two-photon absorption edge of Si
$\omega$/c= 5000~cm$^{-1}$ ($\lambda$=~2000~nm).\cite{Euser05}

A pioneering study of switching 3D photonic materials was done by
Mazurenko \emph{et al.},\cite{Mazurenko03} who reported reflectivity
changes in silica opaline matrices that were infiltrated with Si. This
experiment suffered from several limitations; firstly, the limited
refractive index contrast was insufficient for a band gap to open
up. Secondly, since the experiments were performed at probe
frequencies above the electronic band gap of Si, the transparency of
the unswitched crystal is limited by intrinsic absorption. Moreover,
due to a short Drude damping time for amorphous Si $\tau_{Drude}$=
0.5 fs, the maximum feasible refractive index change is limited by
the amount of induced absorption. Original switching experiments in
Si inverse opals were reported by Becker \emph{et al.} who studied transmission changes.\cite{Beck05} The induced absorption in their crystal was strongly reduced by annealing the Si-backbone, causing $\tau_{Drude}$ to increase drastically from 0.5
fs to 10 fs, resulting in a strong reduction of the induced
absorption.  Unfortunately, however, this study was limited to the frequency range of first order Bragg diffraction where a pseudogap is expected, but not the photonic band gap.

In this article we study ultrafast switching of inverse opal photonic band gap crystals.
There are several reasons why inverse opals are highly suitable for all-optical switching experiments. Firstly, their fabrication is relatively simple, which has allowed inverse opals to be studied extensively. The abundance of
prior static reflectivity experiments helps us to interpret our switching data.\cite{Wijnhoven98,Blanco00,Vos00,Vlasov01,Palacios02} Secondly, the
thickness of opaline crystals is not limited by the fabrication
process, in contrast to crystals that are grown by lithographic
techniques.\cite{Euser06b} Thirdly, band structure calculations for
inverse opals are easily available,\cite{Busch98,Koenderinkthesis}
facilitating the interpretation of the observed stop bands in our
spectra. Fourthly, the crystals can have a sufficiently large refractive
index contrast for a band gap to open up in the range of second
order Bragg diffraction,\cite{Sozuer92} while in the range of first
order Bragg diffraction a pseudo gap occurs.\cite{Ho90} In the
region of the band gap, switching is expected to lead to ultrafast
changes in the density of states.\cite{Johnson02} Finally,
experimental control of the dynamics of spontaneous emission from
quantum dots inside static photonic crystals was recently
demonstrated with inverse opals.\cite{Lodahl04}
Thus, it has been predicted that the spontaneous emission of light sources inside such crystals can indeed be switched on and off.\cite{Johnson02}
We therefore expect that a study of ultrafast switching inverse opals in
the range of second order Bragg diffraction is timely.

\section{EXPERIMENTAL}
\label{experimental}

\subsection{Sample} \label{experimental:samples}

The Si inverse opal photonic crystal was made by infiltrating Si in
a silica opaline template. The template was grown on a Si wafer
substrate by a vertical controlled drying method.\cite{Jiang99} The
resulting 3D silica template extends over 5$\times$5 mm, and was
infiltrated with Si using chemical vapor deposition at 500
$^{\circ}$C.\cite{Vlasov01} Subsequently, the sample was annealed
for one hour at 750 $^{\circ}$C in vacuum. During the annealing
process, the amorphous Si in the structure crystallizes into
poly-Si, as was confirmed by Raman scattering measurements on a
separately deposited thin layer of polycrystalline Si that served as
a reference sample.\cite{Kalkmanthesis, Kalkman04} Finally, the SiO$_2$
template was etched away by a buffered hydrofluoric solution,
resulting in a high-quality 3D air-sphere crystal that is supported
by a poly-Si backbone.

\begin{figure}[!ht]
\begin{center}
\includegraphics[width=0.7\linewidth]{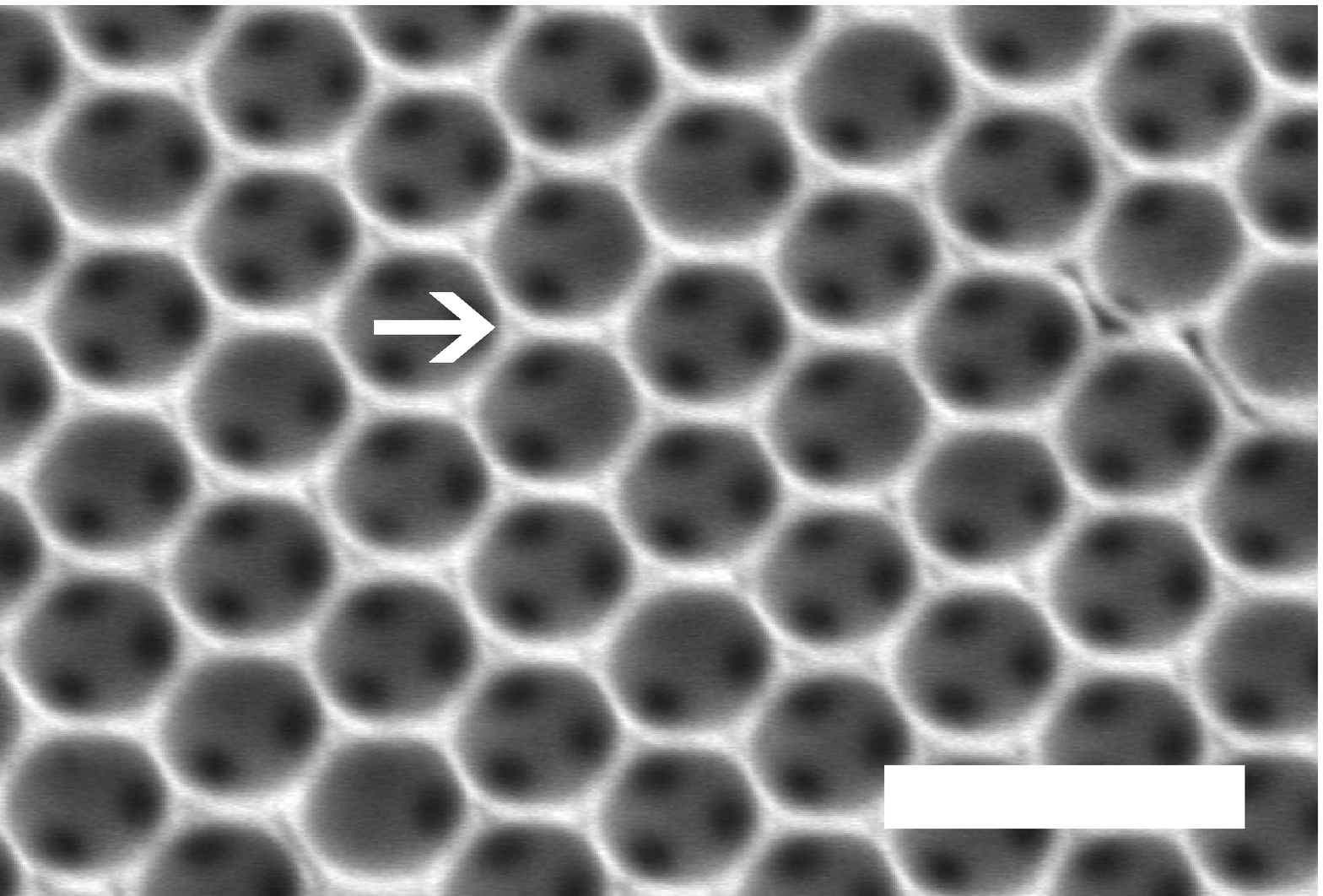}\\
\caption{\label{invopals:fig:detail}High resolution SEM
image of the Si inverse opal after HF etching. The scale
bar is 2 $\mu$m. The arrow indicates an almost filled
interstice in the structure. From this image we estimate
the radius of the windows that interconnect the air spheres
to be $r_{cyl}$= 0.16$\pm$0.05 $r_{sphere}$.}
\end{center}
\end{figure}

We have obtained the microscopic structure of our on-chip Si inverse
opal crystal from high resolution scanning electron micrographs
(SEM). From planar and cross-sectional SEM images we infer that our
inverse opal is a fcc crystal. The size of the lattice parameter of
the fcc-lattice is obtained from Fig.~\ref{invopals:fig:detail},
which shows a top view of a \{111\} plane in the crystal. We find
the lattice parameter to be \emph{a}= 1427$\pm$20 nm, by measuring
the lattice parameter along the three \{$h\bar{h}0$\} in-plane
lattice directions, which allows us to correct for a 22$^\circ$ tilt
of the sample.\cite{Wijnhoven01} From the lattice parameter and the
number of terrace steps, we deduce a sample thickness of 7$\times
d_{111}$= 7$\times$\emph{a}$/\sqrt{3}$= 5.8 $\mu$m.

We also observe from Fig.~\ref{invopals:fig:detail} that the
interstices (indicated by arrows) in the \{111\} surface appear to
be small. To describe the crystal with band structure calculations,
we use a model of close packed air spheres (radius $r_{in}$=
a/$\sqrt{8}$) surrounded by spherical shells (radius $r_{out}$)
connected by cylindrical windows (radius $r_{cyl}$) see
Ref.~\onlinecite{Vos00}. From the almost filled interstices in
Fig.~\ref{invopals:fig:detail}, we estimate an outer
shell radius of $r_{out}$= 1.15$\pm$0.02$r_{in}$. The radius of the air holes
that interconnect the air spheres in the crystal are measured to be
$r_{cyl}$= 0.26$\pm$0.05 $r_{in}$. This method results in a volume fraction of
the solid material of about $\Phi_{Si}$= 21.3 $\%$. Nevertheless, it
should be realized that estimating volume fractions from SEM images
can be problematic, as was found in in-situ x-ray
experiments.\cite{Wijnhoven01}

From the structural parameters obtained, we have calculated the
photonic band structure by a plane-wave expansion method described
in detail in Ref.~\onlinecite{Koenderinkthesis}. We have used a
dielectric constant $\epsilon_{Si}$= 12.74 (at $\omega$= 6535
cm$^{-1}$) that was measured on a separately deposited reference
sample.\cite{Kalkmanthesis} The resulting photonic band structure
calculation is plotted in
Fig.~\ref{invopals:fig:switchedspectrum}(A). The open symbols in
Fig.~\ref{invopals:fig:switchedspectrum}(B) represent a linear
reflectivity spectrum of the sample. It is remarkable that the
measured peaks correspond to the predicted stop gaps, since our
model does not include freely adjustable parameters.

\subsection{Ultrafast switching setup}
\label{experimental:setup}

Our setup consists of a regeneratively amplified Ti:Sapph laser
(Spectra Physics Hurricane) which drives two optical parametric
amplifiers (OPA, Topas). Both OPAs have a continuously tunable
output frequency between 3850 and 21050~cm$^{-1}$, with pulse
durations of 150~fs and a pulse energy $E_{pulse}$ of at least
20~$\rm{\mu}$J. The independent tunability of our OPAs allows us to
optimize the pump frequency, while scanning the probe frequency over
a broad frequency range. The pump beam is incident at $\theta$=
15$^\circ$, and has a much larger Gaussian focus of 133~$\mu$m full width at half maximum (FWHM)
than the probe, providing good lateral spatial homogeneity.  The
probe beam is incident at normal incidence $\theta$= 0$^\circ$, and
is focused to a Gaussian spot of 28~$\mu$m FWHM at a small angular
divergence NA= 0.02. Therefore, we ensure that only the central flat
part of the pump focus is probed. The reflectivity was calibrated by
referencing to a gold mirror. A versatile measurement scheme was
developed to subtract the pump background from the probe signal, and
to compensate for possible pulse-to-pulse variations in the output
of our laser, see Ref.~\onlinecite{Euser06b}.

\section{RESULTS AND DISCUSSION}
\label{results}

\subsection{Linear reflectivity spectra}
\label{invopals:results:linearspectra}

The linear reflectivity spectra in
Fig.~\ref{invopals:fig:switchedspectrum}(B) demonstrates three stops
band in the frequency range $\omega$= 5000-7000 cm$^{-1}$, similar
to earlier work.\cite{Blanco00,Vos00,Vlasov01,Palacios02} We compare
them to a calculated band structure diagram, and label the stop
bands. Stop band I at frequency $\omega$= 5320 cm$^{-1}$ displays a
maximum reflectivity R= 51$\%$ and is identified with the $\Gamma$-L
stop gap at frequency a/$\lambda$= 0.76 in the calculated
bandstructure shown in Fig. \ref{invopals:fig:switchedspectrum}(A).
At $\omega$= 5950 cm$^{-1}$, we observe a stop band, labeled II,
with maximum reflectivity R= 60$\%$. Stopband II corresponds to the
stop gap at a/$\lambda$= 0.85 in
Fig.~\ref{invopals:fig:switchedspectrum}(A). It is important to note
that stop band II overlaps the frequency range of the predicted band
gap that is centred around a/$\lambda$= 0.85. Stop band III at
$\omega$= 6500 cm$^{-1}$ aligns with the stop gap at a/$\lambda$=
0.94. The frequency of the small peak at $\omega$= 7616 cm$^{-1}$
matches that of the stop band at a/$\lambda$= 1.08.

We have systematically reproduced our data on various positions on
the sample. The peak reflectivity of all peaks varies by less than
10$\%$ with position. The variation is possibly due to variations in
the density of lattice defects throughout the crystal. The center
frequency of the stop bands, however, were found to be independent
of position on the sample (within $\Delta\omega/\omega <$ 1$\%$),
which leads to the important conclusion that the crystal lattice is
indeed the same everywhere in the sample.

The agreement between reflectivity peaks and theoretical stop gaps demonstrates
that expected surface effects do not impede the ability of exploiting reflectivity measurements to probe bulk properties. We must note that, despite the apparent good agreement, a direct comparison between reflectivity data and band structure calculations
is strictly speaking not allowed, since band structure calculations
assume an infinitely extended crystal, and do not take into account
surface effects. For instance, a large impedance mismatch of
external modes and internal modes as a result of flat bands in the
band structure can also lead to an increased reflectivity. A stop
gap in the band structure calculations, however, should always
result in an increased reflectivity, which is indeed the case here.

\subsection{Switched reflectivity spectra}
\label{invopals:results:switchedspectra}

\begin{figure}[!ht]
\begin{center}
\includegraphics[width=0.7\linewidth]{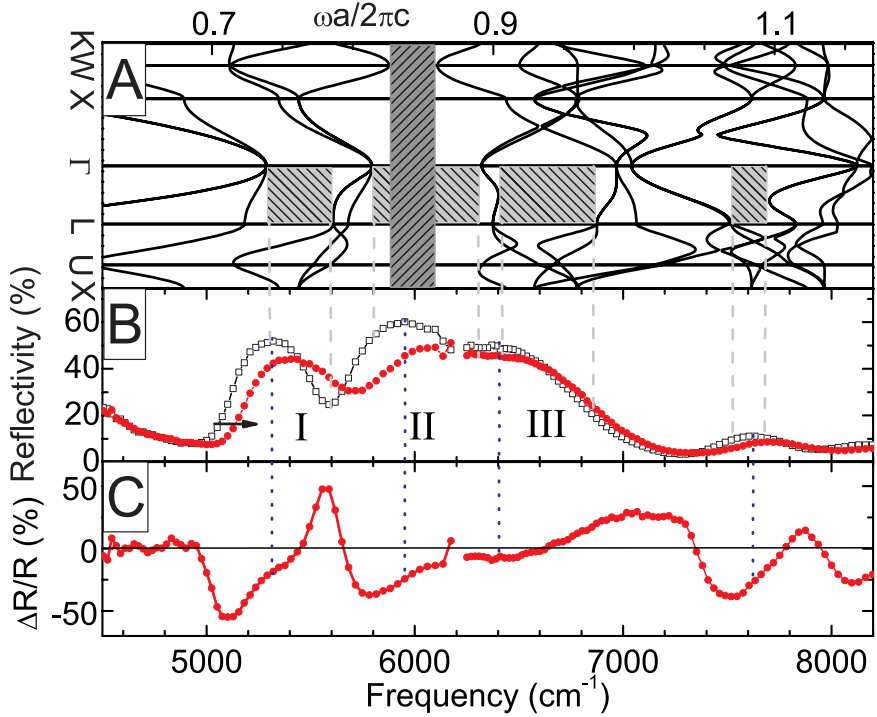}\\
\caption{\label{invopals:fig:switchedspectrum} (A) Photonic band
structures for fcc close packed air spheres (radius r= a/$\sqrt{8}$)
surrounded by spherical Si shells (radius 1.15) connected by
cylindrical windows (radius 0.264r). The volume fraction of solid
material is about $\Phi_{Si}$= 21.3$\%$ ($\epsilon_{Si}$= 12.74).
The frequency scale corresponds to the one in (B) (C) for lattice
parameter a= 1427 nm. The four light gray areas indicate stop gaps
that occur in the $\Gamma$-L direction. The dark gray bar indicates
the frequency range of the band gap. (B) Unswitched (open squares)
and switched (closed circles) reflectivity spectra of the sample at
normal incidence measured with our OPAs. The observed peaks in the frequency
range of second order Bragg diffraction are labeled I, II, and III.
In experiments on the red part of the spectrum
($\omega_{probe}<$6250 cm$^{-1}$), the pump frequency was chosen to
be $\omega_{pump}$= 6450 cm$^{-1}$. In experiments on the blue edge
(6250 cm$^{-1}<\omega_{probe}$), the pump frequency was reduced to
$\omega_{pump}$= 5000 cm$^{-1}$. The pump irradiance for the
switched spectrum was $I_0$= 11$\pm$2 GWcm$^{-2}$ on the red part
and $I_0$= 24$\pm$2 GWcm$^{-2}$ on the blue part of the stop band.
The switched spectra were measured at a pump-probe time delay of
$\approx$1 ps. We observe a large blue shift of up to 1.5$\%$ of the
complete stop band in the range of second order Bragg diffraction as
well as of features outside the stop band. (C) Differential
reflectivity spectrum. A blue shift of peak I at $\omega$= 5320
cm$^{-1}$ is illustrated by strong decreases and increases of the
reflectivity below and above this frequency respectively. Similar
dispersive curves are centered at frequencies of 5960 (II), 6450 (III), and 7600 cm$^{-1}$ (vertical dotted lines).}
\end{center}
\end{figure}

We have induced large and ultrafast reflectivity changes in
our crystal by optically exciting free-carriers. The
ultrafast response of the stop bands is acquired by
measuring the reflectivity spectra at fixed probe delays.
In experiments on the red part of the spectrum
($\omega_{probe}<$6250 cm$^{-1}$), the pump frequency was chosen to
be $\omega_{pump}$= 6450 cm$^{-1}$. In experiments on the blue edge
(6250 cm$^{-1}<\omega_{probe}$), the pump frequency was reduced to
$\omega_{pump}$= 5000 cm$^{-1}$.  The switched reflectivity spectra were
measured in the same run and on the same spot on the sample as the
linear data are shown as closed circles in
Fig.~\ref{invopals:fig:switchedspectrum}(B).  Due to dispersion in
the probe path of the setup, there is a $<$ 500 fs variation of the
delay time with frequency, we therefore measure the switched
reflectivity at a fixed time delay of $\tau$ $\approx$ 1 ps.
Figure~\ref{invopals:fig:switchedspectrum}(B) shows measurements of
the switched and linear reflectivity in the range of second order
Bragg diffraction of the sample at normal incidence.

The dispersion and propagation of light in the crystal is
strongly modified by the switching. Consequently, we
observe large variations in the reflectivity near the peaks
in the region of second order Bragg diffraction. On the red
edge of peak I, near $\omega$= 5110 cm$^{-1}$, the
reflectivity strongly decreases, while on the blue edge, at
$\omega$= 5500 cm$^{-1}$ we observe a strong increase in
the reflectivity, indicative of a blue shift of the entire
peak. Our crystals are therefore highly suitable to control
the directional propagation of light.

The magnitude of the frequency shift of the peaks was obtained by
measuring the frequency position of the red edge of stop band I,
indicated by the black arrow in
Fig.~\ref{invopals:fig:switchedspectrum}(B). The switching moves the
edge towards higher frequencies, while the slope of the stop band
edge remains unchanged. Only at the highest intensity used, induced
absorption can slightly change the slope of the edge. However, the
contribution of this effect to the measured shift is negligibly
small.

The blue-shift of peak I is as large as 80 cm$^{-1}$ or
$\Delta\omega/\omega$= 1.5$\%$. The same effect occurs on
the red edge of peak II, near $\omega$= 5800 cm$^{-1}$,
where the reflectivity decreases, and on the blue edge of
peak III, near $\omega$= 7000 cm$^{-1}$, where the
reflectivity increases. The blue edge of stop band III has
blue-shifted by 50 cm$^{-1}$ or $\Delta\omega/\omega$=
0.7$\%$.  Importantly, all stop bands have shifted towards
higher frequency. We therefore conclude that switching has
reduced the average refractive index of the crystal.

The shift of the stop bands is clearly evidenced by the
dispersive features in the differential reflectivity of the
sample that is plotted in
Fig.~\ref{invopals:fig:switchedspectrum}(C). On the red
edge of stop band I, at $\omega$= 5110 cm$^{-1}$, we
observe a large decrease in the reflectivity by
$\Delta$R/R= -54$\%$, while at $\omega$= 5500 cm$^{-1}$ we
observe a strong increase in the reflectivity by
$\Delta$R/R= 49$\%$. This distinct dispersive shape that is
centered at around 5320 cm$^{-1}$ is related to a large
blue shift of stop band I; the observation of positive differential
reflectivity indicates that the induced absorption remains small. Peak II and III are slightly
broadened by disorder in the sample and thus appear as a
single peak in the spectrum. On the red edge of the
combined peak, at $\omega$= 5800 cm$^{-1}$ the differential
reflectivity is $\Delta$R/R= -35$\%$, while at the blue
edge of the peak, at $\omega$= 7020 cm$^{-1}$ the
differential reflectivity amounts to $\Delta$R/R= +30$\%$; here the dispersive shape also
has a strong positive component, which again signals low induced absorption. The strong
dispersive shape that is centered around 6450 cm$^{-1}$ is related to a large blue shift of the combined stop bands II and III.  The strong dispersive curve that is
centered around $\omega$= 7600 cm$^{-1}$ shows that the
small peak at this frequency also shifts towards higher
frequency.

The observed shift of stop band II towards higher frequency is
particularly interesting, as this stop band is part of the predicted
band gap for inverse opals. We have thus demonstrated switching of a 3D photonic
band gap, which has not been reported before. The switching process is expected to lead to ultrafast changes of
the density of states inside the crystal.\cite{Johnson02}

Remarkably, we observe that both the low and high frequency edge of
the stop bands have shifted. This indicates the absence of separate
dielectric and air bands in the range of second order Bragg
diffraction in inverse opals, which is consistent with predictions
based on quasi-static band structure calculations by
Ref.~\onlinecite{Johnson02}. From our comparison we find a
refractive index change of $\Delta$n'/n'$\simeq$ 2 $\%$ and a
carrier density of $N_{eh}$= 2.1$\times10^{19}$ cm$^{-3}$. The
corresponding plasma frequency is $\omega_p$= 3623 cm$^{-1}$, which
is equal to 0.72$\times \omega_{probe}$. We conclude that excellent
switching conditions indeed appear if the plasma frequency
$\omega_p$ remains smaller than the probe frequency as predicted in
Section \ref{introduction}.


\subsection{Switching time traces}
\label{invopals:results:timetraces}

\begin{figure}[!ht]
\begin{center}
\includegraphics[width=\linewidth]{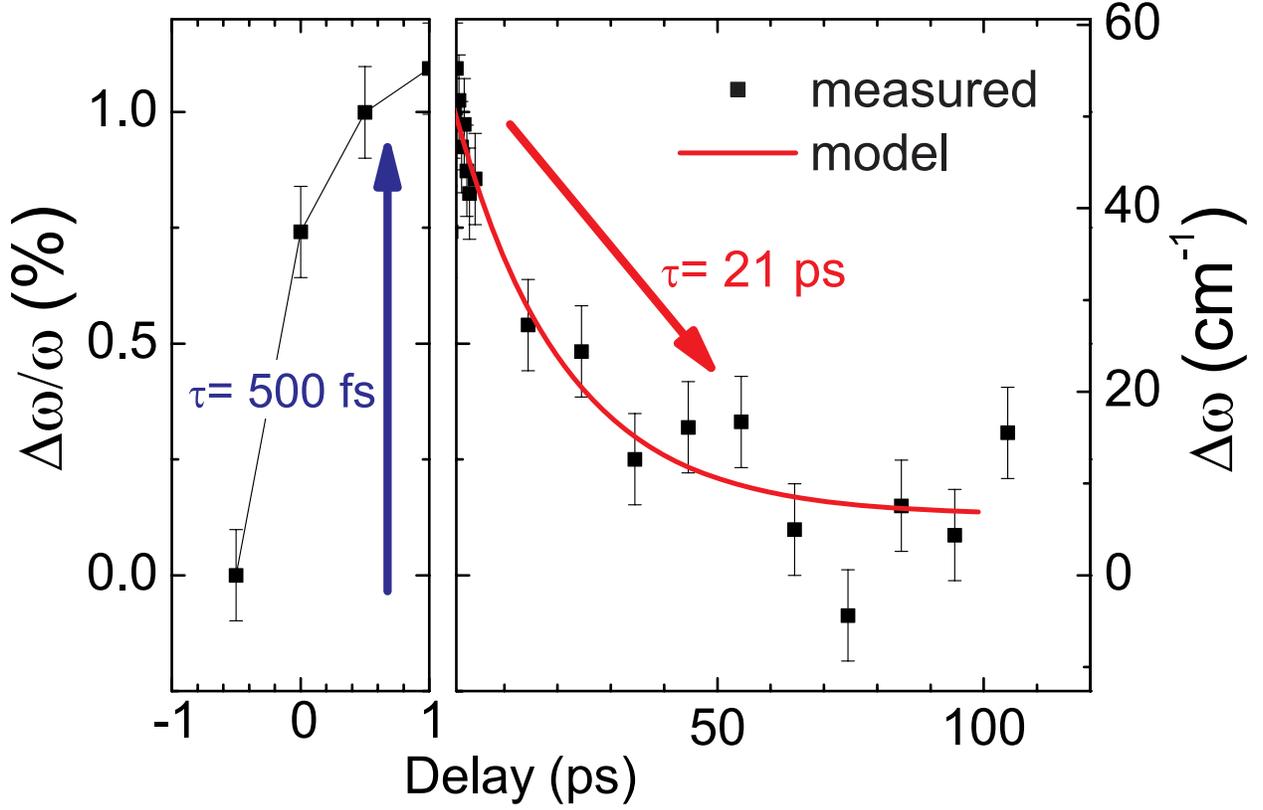}\\
\caption{\label{invopals:fig:shiftvsdelay}Blue shift of the low
frequency edge of stop band I plotted versus probe delay (symbols).
The pump frequency and irradiance were $\omega$= 6450 cm$^{-2}$ and
4$\pm$1 GWcm$^{-2}$ respectively. The large shift amounts to
$\Delta\omega/\omega$= 1.1$\%$ with an exponential growth time of $\tau$= 500 fs (left-hand
panel). The subsequent exponential decay is well fitted with a
single exponential decay $\Delta\omega/\omega$= A+ $\exp(-t/\tau)$
(red curve), where the decay time $\tau$= 21$\pm$4 ps, and the small
offset of $\Delta\omega/\omega$ is A= 0.13$\%$ (right-hand panel).}
\end{center}
\end{figure}

The large and ultrafast shift of the stop band with time is studied
in detail in Fig.~\ref{invopals:fig:shiftvsdelay}. We have measured
the frequency position of the red edge of the stop band at $\omega$=
5045 cm$^{-1}$, at a large range of delay times after excitation.
 From each spectrum, the frequency position of
the low frequency edge was determined at R= 15$\%$
(indicated by the arrow in
Fig.~\ref{invopals:fig:switchedspectrum}(B)). The relative
frequency shift $\Delta\omega/\omega$ is plotted versus
probe delay in Fig.~\ref{invopals:fig:shiftvsdelay}. We
observe a large and ultrafast shift of the stop band edge
from 0 to $\Delta\omega/\omega$= 1.1$\%$ with an exponential growth time of $\tau$= 500 fs,
limited by the pulse duration of our pump pulses.

The effect subsequently decreases exponentially with a decay time of
$\tau$= 21$\pm$4 ps (least squares fit) to a small residual shift
$\Delta\omega/\omega$= 0.1$\%$.  The decay times are much faster
than carrier relaxation times in bulk Si, since our photonic
crystals are made of poly crystalline silicon, whose lattice defects
and grain boundaries act as efficient carrier recombination
traps.\cite{Yu96} The short relaxation time is in good agreement
with the typical carrier relaxation time of 18 ps that we found in
poly-Si woodpile crystals.\cite{Euser06b} The relatively fast decay
time implies that switching could potentially be repeated at GHz
rates, which is relevant to possible future switching and modulation
applications.

\subsection{Induced probe absorption}
\label{invopals:results:inducedabsorption}

\begin{figure}[!ht]
\begin{center}
\includegraphics[width=\linewidth]{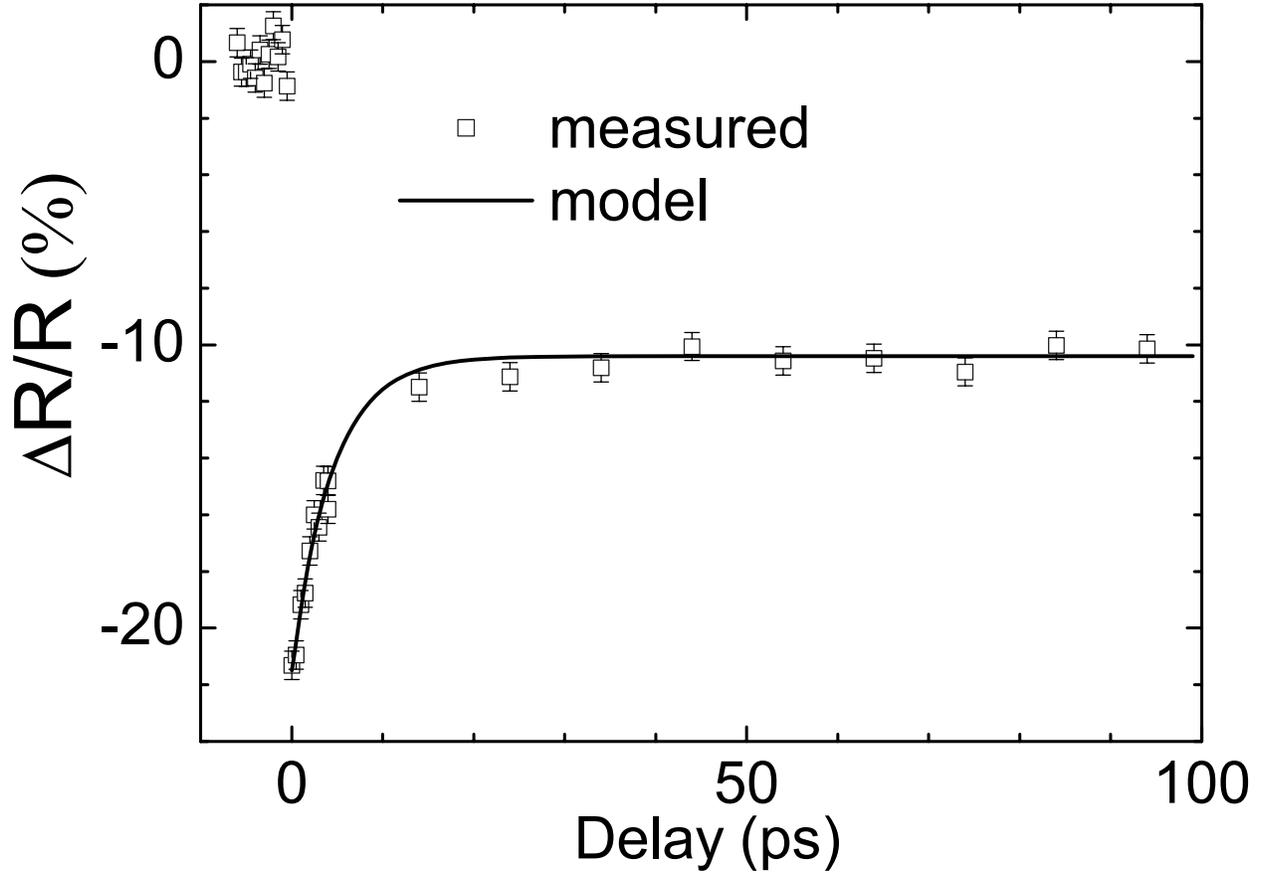}\\
\caption{\label{invopals:fig:drronpeak}Differential
reflectivity at $\omega$= 5882 cm$^{-1}$ versus probe
delay.  The pump frequency and irradiance were $\omega$=
6450 cm$^{-2}$ and 11$\pm$2 GWcm$^{-2}$ respectively. The
large decrease amounts to $\Delta$R/R= -21$\%$ within the
first 500 fs, followed by a decay that is
well fitted with a single exponential $\Delta\omega/\omega$= A+ B$\exp(-t/\tau)$ (curve), with
amplitude B= -11$\%$, decay time $\tau$= 4.5$\pm$0.5 ps and
offset A= -10$\%$.  The offset appears to decay slowly
at ns times and is attributed to the wafer substrate.}
\end{center}
\end{figure}

Besides a frequency shift which is related to a change in the real
part of the refractive index, we observe a decrease of the
reflectivity peak, which is associated with an increase in
$n_{Si}''$. The induced absorption manifests itself in a reduction
of the reflectivity maximum after excitation. We therefore plot in
Fig.~\ref{invopals:fig:drronpeak} the relative decrease in
reflectivity maximum of stop band II at frequency $\omega$= 5882
cm$^{-1}$ as a function of delay time. The data were obtained in the
same run as the experiment shown in
Fig.~\ref{invopals:fig:shiftvsdelay}. The maximum decrease in
reflectivity of the stop band directly after excitation, is
$\Delta$R/R $\approx$ -21$\%$. Note that this decrease is not only
due to absorption but also to a shift. In Ref.~\onlinecite{Euser06b}
we have related a reduction in peak reflectivity of woodpile samples
to the induced absorption through exact model method calculations.
Due to difficulties in calculating the reflectance using a complex
dielectric function in inverse opals, we estimate the induced absorption by comparing
the observed $\Delta$R/R= -21$\%$ to the calculated decrease in Si
woodpile crystals. For woodpile photonic crystals theory that can handle a
complex dielectric function is available.\cite{Euser06b} From this comparison we estimate
an upper bound to the induced absorption in the Si backbone to be $n''_{Si}<0.1$. We use
the Si volume fraction $\Phi_{Si}$= 21.3$\%$, obtained in~\ref{experimental:samples},
to describe our crystal as an effective medium consisting of Si and air. In more advanced
studies, one could also take the spatial distribution of the probe
light in the crystal into account.\cite{Tan04} Since only a fraction
$\Phi_{Si}$ of our crystal absorbs light, we can estimate the
resulting probe absorption length in our switched inverse opal to be
$\ell_{abs}> \lambda/4\pi \Phi_{Si} n''_{Si}$= 6.3 $\mu$m. The
obtained value is larger than the thickness of the sample L=
7$\times d_{\{111\}}$= 5.8 $\mu$m. We conclude that for refractive
index changes larger than 2$\%$, $n''_{Si}$ will increase further,
and consequently the crystal may lose its transparency. Likewise, in applications
in which much smaller changes in the refractive index suffice,
the induced absorption will become negligible small.

Figure~\ref{invopals:fig:drronpeak} also shows how the reflectivity
change evolves in time after the initial decrease to $\Delta R/R=
-21\%$. The effect decays exponentially to $\Delta R/R= -10\%$ with
a decay time of 4 $\pm$ 1 ps, followed by a much slower decay with
about ns decay times.  The fast relaxation process is likely related
to the fast recombination of carriers in the poly-Si backbone of the
crystal that was discussed in the previous Section. It is presently
unclear why the induced absorption decays about five times faster
than the induced stop band shift shown in
Fig.~\ref{invopals:fig:switchedspectrum}(B), which was obtained from
the same set of spectra. The subsequent, much slower recombination
process at ns times is attributed to recombination of excited
carriers in the underlaying Si-wafer substrate. In bulk Si at the
carrier densities in our experiment (10$^{19}$ cm$^{-3}$), carrier
recombination is dominated by Auger effects with recombination times
of the order of 10 ns.\cite{Dziewior77} Any change in the substrate
is likely to change the magnitude of the reflectivity of the whole
sample, while it should not affect the frequency positions of the
stop bands in Fig.~\ref{invopals:fig:switchedspectrum}, which are
related to changes in the backbone of the photonic crystal only.
Indeed, we find that the stop band shift in
Fig.~\ref{invopals:fig:shiftvsdelay} decays rapidly within 100 ps,
to a small offset of a few wavenumbers. Meanwhile, a large part of
the reflectivity decrease in Fig.~\ref{invopals:fig:drronpeak}
is still present after 100 ps, and continues to decay on ns
timescales, consistent with the slow recombination of carriers in
the wafer substrate.

In our experiments we have used two-photon absorption at long pump wavelengths combined with
a large pump focus to maximize the spatial homogeneity of the switched crystals.
The pump fluence in our experiments was typically 25 pJ per unit cell area per pulse. In applications with small active areas, typical for 2D and 3D cavities, spatial homogeneity is not an important issue and thus low pump fluences suffice. Furthermore, the pump fluence can be further reduced by choosing the pump wavelength in the linear absorption range.\cite{Euser05}

\section{CONCLUSIONS}
\label{conclusion}

In this Paper, we have studied all-optical ultrafast switching of a
high-quality 3D Si inverse opal photonic band gap crystal in the
frequency range of second order Bragg diffraction. A spatially
homogeneous free-carrier plasma was optically excited in the crystal
by a two-photon process. We show that for Si inverse opals, good
experimental conditions are found for free-carrier plasma
frequencies around $\omega_p$= 0.3-0.7$\times \omega_{probe}$; large
changes in the refractive index can be achieved, while the crystal
remains transparent after the switching. We find good agreement
between the stop bands in the linear reflectivity spectra and
calculated stop gaps in the frequency range of the band gap.
Switching effects are studied as a function of time delay between
pump and probe pulses. Large ultrafast variations in reflectivity
are observed in the range of second order Bragg diffraction. During
the switching process, all spectral features in the observed stop
bands, shift towards higher frequencies by as much as
$\Delta\omega/\omega$= 1.5$\%$ within a few hundred fs, indicating
the absence of separate dielectric and air bands in our crystal.
From a comparison to quasi-static band structure calculations of
Ref. \onlinecite{Johnson02} we infer a large refractive index change
of about 2$\%$. The deduced refractive index change is predicted to
strongly modify the density of states inside the
crystal.\cite{Johnson02} We have observed a relatively fast decay
time of 21 ps, which implies that switching could potentially be
repeated at a GHz rates, which is relevant to possible future
switching and modulation applications.

\section*{ACKNOWLEDGMENTS}
We thank Cock Harteveld and Rob Kemper for technical support, Martin Wegener,
Ad Lagendijk, Dimitry Mazurenko, and Patrick Johnson for discussions. This work
was also reported on arXiv.org/abs/0705.4250. This work is part of the research program of the "Stichting voor Fundamenteel Onderzoek der Materie"
(FOM), which is supported by the "Nederlandse Organisatie voor Wetenschappelijk Onderzoek" (NWO).
This work was also supported in part by the ACS Petroleum Research Fund, the US NSF (CTS-0332484),
and the US DOE (DE-FG02-06ER46348). We also utilized the Nano Fabrication Center and the
Characterization Facility at the University of Minnesota which receive partial support from
US NSF through the NNIN program.  DJN acknowledges support from the Alexander von Humboldt Foundation.

\newpage

\end{document}